\documentstyle[12pt]{article}

\newcommand{\beq}{\begin{equation}}
\newcommand{\eeq}{\end{equation}}
\newcommand{\bea}{\begin{eqnarray}}
\newcommand{\eea}{\end{eqnarray}}
\newcommand{\id}{\!\!\not\!\partial}

\newcommand{\fs}
{i\kern-.01em\hbox{\raise.25ex\hbox{$/$}\kern-.52em$s$}}

\newcommand{\bs}
{i\kern-.01em\hbox{\raise.25ex\hbox{$/$}\kern-.52em$b$}}
\newcommand{\qs}{/\kern-.52em s}

\newcommand{\D}{{\cal{D}}}

\newcommand{\dd}
{\kern.06em\hbox{\raise.25ex\hbox{$/$}\kern-.40em$\partial$}}

\begin{document}
\title
{Current Algebra and Bosonization in Three Dimensions}
\author{
J.C. Le Guillou$^a$\thanks{Also at {\it Universit\'e de Savoie} and at
{\it Institut Universitaire de France}}\,,
C. N\'u\~nez$^b$ and
F.A. Schaposnik$^a$\thanks{Investigador CICBA, Argentina}
\thanks{On leave from Universidad Nacional de La Plata, Argentina}\\
~
\\
$^a$Laboratoire de Physique Th\'eorique ENSLAPP 
\thanks{URA 1436 du CNRS associ\'ee
\`a l'Ecole Normale Sup\'erieure de Lyon et latex cour\`a 
l'Universit\'e de Savoie}\\
LAPP, B.P. 110, F-74941 Annecy-le-Vieux Cedex, France\\
~\\
$^b$Departamento de F\'\i sica, Universidad Nacional de La Plata\\
C.C. 67, 1900 La Plata, Argentina}
\date{}
\maketitle
\thispagestyle{empty}

%

\begin{abstract}
\normalsize
We consider the fermion-boson mapping in three dimensional space-time,
in the Abelian case,
from the current algebra point of view. We show that 
in a path-integral framework one can derive
a general  bosonization recipe leading, in the
bosonic language, to the correct equal-time
current commutators of the original free fermionic theory.
\end{abstract}

\newpage
\pagenumbering{arabic}

\setcounter{page}{2}
\section{Introduction}

The mapping of 
two-dimensional 
fermionic quantum field theories onto equivalent bosonic
quantum field theories \cite{Lie}-\cite{Wit}, 
commonly called bosonization,
can be viewed as the consequence of a non-trivial
current algebra. In particular, 
difficulties in finding the non-Abelian bosonization recipe
were overcome when the relevance of
current algebra was put forward. Indeed,  the equivalence
of the theory of $N$ massless-free fermions and
the $SO(N)$ Wess-Zumino-Witten model was
proven by showing that the Kac-Moody and Virasoro algebras
coincide in the two theories \cite{Wit}.

Problems encountered
in trying to
extend bosonization to $ d > 2$ dimensions can  be thought as
the result, in part,  of a much more complicated 
algebraic structure of current algebra. In particular,
Schwinger terms
\cite{Got}-\cite{Jac2} which arise when computing
equal time fermion current commutators are divergent
in $d > 2$ dimensions \cite{Bra}-\cite{Ch}, 
in contrast with the finite $d=2$ result. 

In spite of these difficulties and without making appeal to current 
algebra,
a fermion-boson mapping has been  established in $ d=3$ dimensions 
\cite{Lut}-\cite{BFO}. The resulting
bosonization recipes for fermion currents,
valid in different regimes, 
either for free or interacting models, both in the Abelian and 
non-Abelian
cases are the natural extension of the $d=2$ ones \cite{FS}-\cite{FAS}.
However, a detailed analysis of the 
relation between fermionic and bosonic current algebra 
is lacking. We shall address to this
issue in the present work.

As we shall see, one can establish using the path-integral approach
a fermion-boson mapping leading to an {\it exact} bosonization recipe
for fermion currents in the Abelian case. This is due to the fact,  
first exploited
in \cite{JLG} for the $d=2$ case, that external-source dependent terms 
can be factored out very simply within the path-integral approach.
In particular, in $3$ dimensional (Euclidean) space-time, 
in the Abelian case,  the fermion current
bosonizes according to \cite{FAS}
\beq
j_\mu = - i\bar\psi \gamma_{\mu} \psi  \to 
\frac{1}{{\sqrt {4 \pi}}} \epsilon_{\mu \nu\alpha}
\partial_{\nu}A_{\alpha} ~,
\label{aq1}
\eeq
where $A_{\alpha}$ is a vector field. Now, the
exact bosonic Lagrangian equivalent to the
original fermionic one is far too complicated 
to perform exact calculations and it is for this
fact that bosonization implies some approximation in
$d = 3$ dimensions. 
A possible simplification consists in either
neglecting  non-quadratic terms 
in the construction of the bosonic counterpart of a massless
Dirac fermion Lagrangian \cite{Mar}, \cite{BFO} 
or, alternatively,
to perform a low energy approximation  at some stage of the 
derivation so that the results are only valid for {\it very massive}
fermions \cite{FS},\cite{FAS}.
While in the former case the resulting effective action  looks
rather complicated, in the latter  one ends with a very simple
Maxwell-Chern-Simons  (MCS) theory.

MCS theory is very attractive  since
it exhibits plenty of interesting properties \cite{Jac} 
to be welcome both in Particle Physics and in Condensed
Matter Physics \cite{Fradb}, its connection with
$3$-dimensional fermionic theories being precisely of interest
in this last domain. One should remind however that
the equivalence between the free fermion Lagrangian and
MCS theories is not exact
but only valid in the large fermion-mass region, this 
meaning the large-distances regime for fermion fields.
Since current commutators test the short-distance regime, one
should not take the MCS gauge-invariant algebra as a starting
point to reproduce the fermion current commutators.
Indeed,
the commutator algebra obeyed by the electric and magnetic
fields in MCS theories
has been shown to have {\it finite} derivative  terms
\cite{Jac},\cite{DJ}. Were one naively  to use the bosonization 
recipe (\ref{aq1}) in a 
calculation of current commutators,
taking the MCS action as the bosonized equivalent
to that for free fermions,
the
correct {\it divergent} Schwinger terms 
 would be absent. 

In view of the discussion above, the  study of
three dimensional equal-time fermion current commutators 
within the bosonization approach needs
an approximation scheme  
valid for an arbitrary (not necessarily large)
fermion mass.   
To handle this problem, we study in the present work
a quadratic
approximation of the free massive fermion theory following 
a recent proposal \cite{BFO}. In this way, we show
how the correct equal-time current commutators arise in the bosonic 
theory,
coinciding with those calculated in the original free fermion model
\cite{Bra}-\cite{Ch},\cite{dVG}. As it is the case in two dimensions, 
by showing
that the current algebra resulting from the bosonized
theory coincides with that of the original
fermion model, we go a step further in the obtention of
a fermion-boson mapping in $3$ dimensions.

The paper is organized as follows. We give in Section 2 a brief
review of our  bosonization approach for $d \geq 2$ dimensions 
and then discuss in Section 3
how the equivalent bosonic theory can be approximated to
evaluate current commutators in $d=3$. Following the 
Bjorken-Johnson-Low 
method we then show  that the current commutators
obtained according to our bosonization recipe coincide with the 
original
ones arising in the fermionic model. We finally summarize our results
in Section 4. 

\section{Bosonization}
We shall describe in this section how bosonization rule (\ref{aq1}) can
be derived within the path-integral framework. As first stressed
in \cite{JLG}, one important ingredient 
in path-integral bosonization is the introduction of sources which
allow to identify the correct bosonic fields and exhibit the
{\it exact} bosonization recipe for fermion currents. A second
ingredient is the introduction of an auxiliary field
which allows for obtaining in a very simple way the bosonized action.
This approach to bosonization was originally 
introduced in \cite{Bur1} to study two-dimensional bosonization
and then extended to $ d > 2 $ dimensions, both for 
Abelian and non-Abelian cases
\cite{Bur}-\cite{FAS}. We here follow the presentation
introduced in \cite{FAS} but emphasizing on the exactness of the 
bosonization recipe for fermion currents achieved within this approach. 

Consider for definiteness the action for a free massive Dirac fermion,
coupled to an external source $s_\mu$
\beq
{\cal L}_{F}= \bar\psi (\id + m + \fs) \psi
\label{L}
\eeq

The corresponding partition function in $d$ dimensional Euclidean
space-time reads
\beq
Z_{fer}[s] = \int D\bar{\psi}D\psi \exp\left[-\int d^{d}x 
\bar{\psi}(\id  
+ m + \fs)\psi\right] ~.
\label{ferm0}
\eeq
We now perform the following change of fermion variables
\[
\psi =g(x) \psi'
\]
\beq
\bar \psi = \bar \psi' g^{-1}(x) .
\label{chas2}
\eeq
Here $ g(x) = \exp[i\theta(x)] $. Although we shall consider 
just the case of one Dirac fermion and then
 $g \in U(1)$, the non-Abelian
case can be analogously treated.
One can always define for Dirac fermions a path-integral 
measure invariant under
transformation (\ref{chas2}). Then, after the change of variables
the partition function becomes
\beq
Z_{fer}[s] = \int \D\bar\psi \D\psi \exp[
-\int d^dx \bar\psi (\id + m + g^{-1}\id g + \fs) \psi ].
\label{n2}
\eeq
(we have omitted primes in the new fermionic variables). 
It is evident that $i g^{-1}\id g$  in (\ref{n2}) can be
thought as a flat connection and can be then replaced by a
non-trivial gauge field connection $b_\mu$
provided a constraint 
to ensure its flatness is introduced. Then,  
we can rewrite (\ref{n2}) in the form
\beq
Z_{fer}[s] = \int \D \bar\psi \D\psi \D b_\mu 
\delta[\epsilon_{{\mu_{1}} {\mu_{2}}...{\mu_{d}}}
f_{{\mu_{1}}{\mu_{2}}}]
\exp[-\int d^dx \bar\psi (\id + m + \bs + \fs) \psi ].
\label{ntrade}
\eeq
where
\beq
f_{\mu \alpha} = \partial_\mu b_\alpha - \partial_\alpha b_\mu 
\label{nf}
\eeq
The next step in our derivation is to integrate out fermions so that
the partition function becomes
\beq
Z_{fer}[s] = \int Db_\mu  det(\id + m + \bs + \fs)
\delta[\epsilon_{{\mu_{1}} 
{\mu_{2}}...{\mu_{d}}}f_{{\mu_{1}}{\mu_{2}}}] .
\label{in}
\eeq
Now, a trivial shift $ b + s \to b$ in the integration variable $b$ 
puts
the source dependence into the constraint
\beq
Z_{fer}[s]= \int Db_\mu  det(\id + m + \bs)
\delta[\epsilon_{{\mu_{1}} {\mu_{2}}...{\mu_{d}}}
(f_{{\mu_{1}}{\mu_{2}}} - 2\partial_{\mu_{1}} s_{\mu_{2}})]
\label{in22}
\eeq
It is at this point that the bosonic field, equivalent to the original
fermion field, enters into play. Indeed, in order to handle
the flatness condition, one represents the delta function
in (\ref{in22}) in the form
\begin{eqnarray}
& &\delta[\epsilon_{{\mu_{1}} {\mu_{2}}...{\mu_{d}}}
\left(f_{{\mu_{1}}{\mu_{2}}} - 2 \partial_{\mu_{1}} s_{\mu_2} ) \right) 
]
\nonumber\\
& & = \int D\phi_{} \exp \left[ \frac{i}{2}
\int d^dx \epsilon_{{\mu_{1}} {\mu_{2}}...{\mu_{d}}}
\phi_{{\mu_{3}}...{\mu_{d}}}(f_{{\mu_{1}}{\mu_{2}}} - 
2 \partial_{\mu_{1}} s_{\mu_2} ) \right] 
\label{lag22a}
\end{eqnarray}
It is the Lagrange multiplier $\phi_{{\mu_{3}}...{\mu_{d}}}$,
which can be seen as a Kalb-Ramond field \cite{Bur}, that
becomes the bosonic equivalent of the original fermion.  Hence,
the fermionic partition function can be finally written as
\begin{eqnarray}
Z_{fer}[s]  & = &\int D\phi
\exp \left( -i
\int d^dx \epsilon_{{\mu_{1}} {\mu_{2}}...{\mu_{d}}}
(\partial_{\mu_{2}}\phi_{{\mu_{3}}...{\mu_{d}}} )s_{\mu_1} \right) 
\exp\left(-S_{bos}[\phi]\right) \nonumber \\
&\equiv & Z_{bos}[s]
\label{efe}
\end{eqnarray}
with
\beq
\exp\left(-S_{bos}[\phi]\right) = 
\int Db_\mu det(\id + m + \bs) \exp\left( \frac{i}{2}
\int d^dx \epsilon_{{\mu_{1}} {\mu_{2}}...{\mu_{d}}}
\phi_{{\mu_{3}}...{\mu_{d}}}
f_{{\mu_{1}}{\mu_{2}}} \right)
\label{in22f}
\eeq
The central point in our bosonization route is now at sight:
the external source $s_\mu$ has factored out from the auxiliary
field integration so that by simple differentiation
one has the exact bosonization recipe for fermion currents
\beq
\bar \psi \gamma_{\mu_{1}}  \psi \to 
\epsilon_{{\mu_{1}} {\mu_{2}}...{\mu_{d}}}\partial_{\mu_{2}}
\phi_{{\mu_{3}}...{\mu_{d}}} ~.
\label{lance}
\eeq
This, together with the recipe for bosonizing the fermion action
\beq
\int  
\bar{\psi}(\id +  m )\psi d^{3}x \to S_{bos}[\phi]
\label{agre}
\eeq
with $S_{bos}$ defined by eq.(\ref{in22f}) completes
the fermion-boson mapping in the Abelian case.
Of course, one has still to perform the $b_\mu$ integration
to give an explicit expression
for $S_{bos}$.
Whether one would arrive
to a close result for the bosonized action or to an approximate one
will depend
on the possibility of computing the fermion determinant exactly
and then performing the integration over $b$. This can be of course
done in $2$ dimensions where the well-known bosonization recipe can
be derived within this approach in a very simple way \cite{FAS}.
If this were also possible
for $d \geq 2$, then the fermionic degrees of freedom, which 
disappeared
from the partition function, would  be replaced by new bosonic degrees 
of freedom $\phi_{{\mu_{3}}...{\mu_{d}}}$,
in an exactly equivalent bosonic model. Even if one is not able
to compute the fermion determinant exactly (as one does
for $ d = 2$), it is important to note that the
bosonization recipe for the fermion current is exact. It is 
the bosonic action that accompanies this recipe that
has to be approximated.

\section{Current commutators}
We have seen in the precedent
section that starting from the free fermion generating  functional 
in $d$ dimensional space-time in the presence of a source
$s_\mu$, one can systematically construct a bosonic theory 
and obtain in a very
simple way the bosonization recipe for the fermionic current. We shall
specialize in this section to $3$-dimensional
(Euclidean) space, so that the original
fermionic generating functional.

\beq
Z_{fer}[s] = \int D\bar{\psi}D\psi \exp\left[-\int d^{3}x 
\bar{\psi}(\id + m +
\fs  )\psi\right] ~,
\label{ferm}
\eeq
becomes, after bosonization, 
\begin{eqnarray}
Z_{fer}[s] &=& \int DA_\mu \exp(- i  \epsilon_{\mu \nu \alpha}\int d^3x
s_\mu \partial_\nu A_\alpha) \times  \nonumber\\
&&\int Db_\mu det(\id + m + \bs)
\exp(+i \epsilon_{\mu \nu \alpha}\int d^3x
b_\mu \partial_\nu A_\alpha) ~,
\label{bos}
\end{eqnarray}
where  $A_\mu$ is the bosonic field 
($A_\mu = \phi_\mu$) replacing the original fermion field
and 
$b_\mu$ is the auxiliary field to be integrated out.

As discussed in the precedent section,  the
possibility of going on with the bosonization procedure depends on 
one's
ability to evaluate the fermion determinant in the r.h.s. of 
eq.(\ref{bos})
and then integrate out the auxiliary field $b_\mu$. In
three dimensional space-time, one possibility is to
use the results of the $1/m$ expansion \cite{Red} to give an 
expression for
the fermion determinant which will then be valid only for large masses.
Alternatively, one can make an
expansion in powers of $b_\mu$ retaining up to quadratic terms. 
Following this last procedure, 
integrating out $b_\mu$ and diagonalizing the resulting
bosonic action, one ends with \cite{BFO}
\beq
Z_{fer}[s] = Z_{bos}[s],
\label{igu}
\eeq
with
\begin{eqnarray}
Z_{bos}[s] &=&  \int DA_{\mu} \exp \left( -\int d^3x
[\frac{1}{4}F_{\mu\nu}
C_{1}(-\partial^2)F_{\mu\nu}  \right.\nonumber\\
& - &\left.\frac{i}{2}A_{\mu}C_{2}(-\partial^2)
\epsilon_{\mu\nu\lambda}
\partial_{\nu}A_{\lambda}  \right. \nonumber \\
& + &\left. i\ \frac{u_{+} - u_{-}}{2} s_{\mu}
\frac{1}{\sqrt{-\partial^2}}
\partial_{\nu}
F_{\mu\nu}
+ i \frac{u_{+} + u_{-}}{2} s_{\mu}
\epsilon_{\mu\nu\lambda}\partial_{\nu}
A_{\lambda}]\right) 
\label{bos2}
\end{eqnarray}
Here
$C_1$ and $C_2$ are defined as
\beq
C_{1} = 
\frac{1}{2} \frac{|u_{+}|^2 (F - iG) + 
|u_{-}|^2 (F + iG)}{-\partial^{2} F^2 + G^2}
\label{rosa}
\eeq
\beq
C_{2} =\frac{i}{2} 
\frac{|u_{+}|^2 (F - iG) - |u_{-}|^2 (F + iG)}{-\partial^{2} F^2 + G^2}
\eeq
with
$F(-\partial^2)$ and $G(-\partial^2)$ given through their 
momentum-space
representations ${\tilde F}$
and ${\tilde G}$ as \cite{BFO}
\beq
{\tilde F} (k) \;=\; \frac{\mid m \mid}{4 \pi k^2} \,
\left[ 1 
- \displaystyle{\frac{1 \,-\,\displaystyle{\frac{k^2}{4 m^2}}}{(
\displaystyle{\frac{k^2}{4 m^2}})^{\frac{1}{2}}}} \, \arcsin(1\,+
\, \frac{4 m^2}{k^2})^{-\frac{1}{2}} \right] \;,
\label{1.10}
\eeq
\beq
{\tilde G} (k) \;=\; \frac{q}{4 \pi} \,+\, \frac{m}{2 \pi \mid k \mid}
\, \arcsin (1 \, + \, \frac{4 m^2}{k^2} )^{- \frac{1}{2}} \;,
\label{1.11}
\eeq
Quantities $u_\pm$ are arbitrary functions of the momentum, 
arising in the
diagonalization procedure. Concerning $q$, it is a regularization 
dependent
parameter which can assume any integer value.
Current correlation functions and, 
{\it a fortiori}, current commutators  can be proven to
be $u$-independent. For simplicity we  shall take them equal and 
constant,
 $u_\pm = u $ 
so that eq.(\ref{bos2}) reduces to
\begin{eqnarray}
Z_{bos} &=& \int DA_{\mu} \exp \left[-\int d^3x (\frac{1}{4}F_{\mu\nu}
C_{1}F_{\mu\nu} - \frac{i}{2}A_{\mu}C_{2}\epsilon_{\mu\nu\lambda}
\partial_{\nu}A_{\lambda}  \right. \nonumber\\
&  & \left. + {i}{u} s_{\mu}
\epsilon_{\mu\nu\lambda}\partial_{\nu}A_{\lambda}) \right]
\label{bos3}
\end{eqnarray}
with $C_1$ and $C_2$ now given through their momentum-space
representation ${\tilde C}_1$ and ${\tilde C}_2$
\beq
{\tilde C}_{1}(k) = \vert u \vert^2
\frac{{\tilde F}(k)}{k^{2} {\tilde F}^2(k) + {\tilde G}^2(k)}
\label{c1}
\eeq
\beq
{\tilde C}_{2}(k) = \vert u \vert^2\frac{{\tilde G}(k)}{k^{2}{\tilde 
F}^2(k) + 
{\tilde G}^2(k)}
\label{c2}
\eeq 

Equation (\ref{bos3}) gives the  bosonized version of 
the original fermion model partition function in the presence
of an external source, in the quadratic approximation. We 
again see that 
the bosonization recipe advanced in \cite{FS} for the large
mass case is in fact valid at all scales,
\beq
 \bar \psi \gamma_\mu \psi \to  
\frac{i}{\sqrt{4\pi}} \epsilon_{\mu \nu \alpha}\partial_\nu A_\alpha
\label{curm}
\eeq
(To compare with ref. \cite{FS} we have made the choice 
$u = i/\sqrt{4\pi}$).
The point is that the r.h.s. in eq.(\ref{curm})
involves a bosonic field with dynamics
governed by the  bosonic 
action 
\beq
S_{bos} = \int[\frac{1}{4}F_{\mu\nu}
C_{1}(-\partial^2)F_{\mu\nu} - \frac{i}{2}A_{\mu}C_{2}(-\partial^2)
\epsilon_{\mu\nu\lambda}
\partial_{\nu}A_{\lambda}]d^3x  
\label{ac3}
\eeq
As discussed in detail in refs.\cite{Bur} and \cite{FAS}, in
the path-integral approach to bosonization, the bosonic field
naturally appears  as a vector field with a gauge invariant
action (\ref{ac3}) so that the corresponding partition function
$Z_{bos}$ will require a gauge fixing.

One can easily see that eqs.(\ref{bos3})-(\ref{ac3}) reproduce, 
in the large mass limit, the bosonization result given
in \cite{FS},\cite{FAS}. Indeed, 
taking the limit of large masses in the expressions for
$C_1$ and $C_2$ (eqs.(\ref{c1})-(\ref{c2})) one reobtains
the derivative expansion  result for the 3-dimensional
fermion determinant \cite{Red}-\cite{DesR} so that
the  generating functional 
(\ref{bos3}) becomes within this approximation \cite{BFO}
\begin{eqnarray}
Z_{bos}[s] & &\simeq 
\int DA_\mu \exp\left[ - 
\int d^3x (\mp \frac{i}{2}\epsilon_{\mu \nu \alpha} 
A_\mu \partial_\nu A_\alpha 
+\frac{1}{12\vert m\vert} F_{\mu \nu}^2 \right.
 \nonumber\\
&  &\left. -\frac{1}{\sqrt{4\pi}}\epsilon_{\mu \nu \alpha}
s_\mu \partial_\nu A_\alpha)\right]
\label{mgran}
\end{eqnarray}
(Again, to compare with \cite{FS}
we have made $u = i/(\sqrt{4\pi})$ and chosen the
arbitrary parameter $q$ so that $q \pm m/\vert m \vert= \pm 1$.)
From 
eq.(\ref{mgran}) we see that
the fermion system is described for large masses
by a Maxwell-Chern-Simons theory.
Now, the gauge invariant algebra of  
such theory has been studied in refs.\cite{Jac},\cite{DJ}. 
One has for instance in our case,
\beq
[E_i({\vec x},t),B({\vec y},t)] = 
 - 3  \, \vert m \vert \epsilon_{ij}
 \partial_j\delta^{(2)}({\vec x} - {\vec y}) 
\label{c21}
\eeq
If one now 
relates the electric field $E_i = F_{i0}$ and the magnetic field 
$B = \epsilon_{ij}\partial_iA_j$  to the fermionic
currents through the bosonization recipe (\ref{aq1}),
\beq
j_o \to \frac{1}{\sqrt{4\pi}} B
\label{b1}
\eeq
\beq 
j_i \to \frac{1}{\sqrt{4\pi}}
 \epsilon_{ij}E_j
\label{b2}
\eeq
then, the resulting fermion current commutator
algebra is not
the 
one to be expected for three-dimensional free fermions. Indeed, 
the $d=3$ fermion current algebra should contain an infinite 
Schwinger term 
\cite{BJ}-\cite{dVG} which is
absent in eq.(\ref{c21}). 
The point is that eq.(\ref{mgran})
ensuring that the resulting bosonic theory
is a Maxwell-Chern-Simons theory,
is valid only for large fermion mass while calculation of
equal-time current commutators imply, as we shall see,
a limiting procedure which cannot be naively followed
for large masses.  

Since the exact bosonic partition function
is much too complicated to
handle, a possible strategy is to  
use the quadratic (in auxiliary fields)
approximation for fermions with an
arbitrary (not necessarily large) mass so as to
obtain a bosonized version of the original fermionic
model in which the equal-time limit can be safely taken. 
One should then compute current commutators for this bosonized theory,
and test whether they coincide with those satisfied 
by fermionic currents in the original model.
As we shall see, 
our bosonization scheme does
reproduce the correct equal time commutator algebra; moreover,
although we find equal time commutators for arbitrary mass,
it becomes apparent that the divergent Schwinger term
can already be obtained in the small mass limit, where the
bosonized action takes a reasonably simple form so that finally
one disposes of a simple bosonization recipe covering
both the small and large distances regimes. The recipe would
be unique in the sense it gives a unique relation between the
fermionic current and the bosonic field curl although the
effective bosonic theory would differ in 
each regime.

To proceed according to this strategy, let us rewrite 
the partition function (\ref{bos3})
(taking from here on $u = i/\sqrt{4\pi}$)
in the form
\begin{eqnarray}
Z_{bos}[s] &=& \int DA_{\mu} \exp 
\left[-\frac{1}{2}\int d^{3}x d^3y 
A_{\mu}(x) D_{\mu\nu}(x,y) A_{\nu}(y)
 \right. \nonumber\\
& &\left. - \frac{1}{\sqrt{4\pi}} 
\int d^3x A_{\lambda}\partial_{\nu} s_{\mu} \epsilon_{\mu\nu\lambda}
 \right]
 \label{lata}
\end{eqnarray}
or
\beq
Z_{bos}[s] = \left[ det D_{\mu\nu}\right]^{- \frac{1}{2}} 
\exp\left[\frac{1}{8\pi}
\int d^3x d^3y\partial_{\nu} s_{\mu}(x)\epsilon_{\mu\nu\lambda} 
D_{\lambda\rho}^{-1}(x,y)\partial_{\sigma} s_{\tau}(y)
\epsilon_{\rho \sigma \tau}\right]
\label{integ}
\eeq
where 
$D_{\mu\nu}^{-1}$ is just the propagator of the bosonic action,
which in the Lorentz gauge we adopt from here on reads
\beq
D_{\mu\nu}^{-1}(x,y) = \int \frac{d^{3}k}{{(2\pi})^3}  
\left[P(k) g_{\mu\nu}  + Q(k) k_\mu k_\nu
+ R(k) \epsilon_{\mu\nu\alpha}k_{\alpha}  
\right]\exp{i k(x - y)}
\label{funn}
\eeq
with
\beq
P(k) = \frac{{\tilde C}_1(k)}{k^2 {\tilde C}_1^2(k) + 
{\tilde C}_2^2(k)}
= 4\pi {\tilde F}(k)
\label{P}
\eeq
\beq
Q(k) = \frac{{\tilde C}_1(k)}{k^2 {\tilde C}_1^2(k) +
 {\tilde C}_2^2(k)} 
\left(\frac{{\tilde C}_2(k)}{k^2{\tilde C}
_1(k)}\right)^2
\label{Q}
\eeq
\beq
R(k) = \frac{{\tilde C}_2(k)}{k^2(k^2 {\tilde C}_1^2(k) + 
{\tilde C}_2^2(k))}
\label{R}
\eeq

Let us
briefly recall how one can compute current commutators
within the path-integral scheme using the so-called
BJL method \cite{BJL}-\cite{BJL2}, \cite{Jac2}. To this end we define
the correlator
\beq
G_{\mu\nu}(x,y) = \left. \frac{\delta^{2} log Z_{bos}[s]}{\delta
 s_{\mu}(x) 
\delta s_{\nu}(y)}\right|_{s=0} 
\label{G}
\eeq 
from which one can easily derive equal time current commutators
using the relation 
\beq
<[j_0({\vec x},t),j_i({\vec y},t)]> = 
\lim_{\epsilon \to 0^+}
[G_{0i}(\vec{x},t + \epsilon;{\vec y},t) - 
G_{0i}(\vec{x}, t-\epsilon; {\vec y},t)]
\label{conmut}
\eeq
The current commutator evaluated using eqs.(\ref{G})-(\ref{conmut})
corresponds to the {\it bosonic} partition function $Z_{bos}[s]$ 
obtained in eq.(\ref{bos3}). That is, eq.(\ref{conmut}) gives the
equal-time commutator for the bosonic currents 
$j_\mu = (1/\sqrt{4\pi}) \epsilon_{\mu \nu \alpha} \partial _\nu 
A_\alpha $. This result should 
then be
compared with that arising in the original $3$-dimensional fermionic
model for which $j_\mu = - i\bar \psi \gamma_\mu \psi$ \cite{dVG}.

Starting from eqs.(\ref{integ})-(\ref{funn}) and using the
BJL method we get, after some calculations,
\beq
G_{\mu \nu}(x,y) = 
-\frac{1}{4 \pi} \epsilon_{\mu \alpha \rho} \epsilon_{\nu \beta \sigma}
\partial_\alpha \partial_\beta D^{-1}_{\rho \sigma}
\label{ag1}
\eeq
or
\beq
G_{\mu \nu}(x,y) = \frac{1}{4 \pi}
\int \frac{d^3k}{(2\pi)^3} [P(k)(k^2 g_{\mu \nu} - k_\mu k_\nu) 
+k^2 R(k) \epsilon_{\mu \nu \alpha} k_\alpha]\exp[ik(x-y)]
\label{ag2}
\eeq
With this, we can rewrite eq.(\ref{conmut}) in the form
\beq
<[j_0({\vec x},t),j_i({\vec y},t]> = 
\lim_{\epsilon \to 0^+} I^\epsilon({\vec x} - {\vec y})
\label{ep}
\eeq
with
\beq                               
I^\epsilon({\vec x}) = -2i \int \frac{d^{3}k}{({2\pi})^3} 
k_{0}k_{i}
{\sin(k_0 \epsilon)} {\tilde F}(k)
\exp{i \vec{k}.\vec{x}}
\label{int1}
\eeq
where we have written $(k_\mu) = (k_o,k_i)$, $i=1, 2$. It will
be convenient to define 
\beq
k'_0 = \epsilon k_0
\label{ko}
\eeq
In terms of this new variable and using the explicit form for 
${\tilde F}(k)$
given by eq.(\ref{1.10}), with $k = {(k_0^2 + \vec k^2)}^{1/2}$,
integral $I^\epsilon$ becomes
\beq
I^\epsilon({\vec x}) = - \frac{1}{8 \pi^2 \vert m \vert} 
\frac{1}{\epsilon^2} \partial_i \int \frac{d^2k}{(2\pi)^2}
\exp{i \vec{k}.\vec{x}} \int_0^\infty  dk'_0 k'_0 \sin k'_0 f(y)
\label{uf1}
\eeq
where
\beq
{f} (y) \;=\; \frac{1}{y} \,
\left[ 1 - \frac{(1-y)}{\sqrt{y}}
 \arcsin \frac{1}{\sqrt{1 + (1/y)}} \right]
\label{1.101}
\eeq
and we have defined
\beq
y = \frac{k^2}{4m^2} = 
\frac{{k'}_0^2 + \epsilon^2 {\vec k}^2}{4 \epsilon^2 m^2}
\label{nues}
\eeq
One can now see that  $ y \to \infty$ for 
$\epsilon \to 0$ and fixed $m$. Then, expanding in powers of $1/y$
one has $f(y) \sim \pi/(2\sqrt y)$
and then using distribution theory to define the integral
over $k'_0$ one finds
\beq
<[j_0({\vec x},t),j_i({\vec y},t]>  =  -\frac{1}{8 \pi}
\lim_{\epsilon \to 0} \frac{1}{ \epsilon}
\partial_i\delta^{(2)}(\vec x - \vec y)  
\label{concop}
\eeq
This result for the
equal-time current commutator, evaluated
within the bosonized theory,
shows exactly the infinite Schwinger term that is found,
using the BJL method, for free
fermions in $d=3$ dimensions \cite{dVG}. As it happens in $d=4$ 
dimensions
\cite{Bra}-\cite{Ch}, we see
from eq.(\ref{concop}) that the commutator at {\it unequal} times
is well defined: divergencies appear only when one takes the
equal-time limit. 

It is interesting to evaluate the  next order
vanishing in the equal-time limit 
so as to compare with
the result for the original fermion model reported in the
literature \cite{dVG}.
Due to the combination
$\epsilon m$ appearing in $y$ 
one can make $m \to 0$ in the equal-time limit
without loss of generality. 
Then, 
one can first consider the limit of small mass, expand $f$,
integrate exactly over $k'_0$ and
then take  $\epsilon \to 0$. Now, for small mass one has
\beq
f(y) = \pi \vert m\vert( \frac{1}{({\vec k}^2 + ({k'_0}/
\epsilon)^2)^{1/2}}
- 4  m^2 \frac{1}{({\vec k}^2 + ({k'_0}/\epsilon)^2)^{3/2}} + \ldots )
\label{exp7}
\eeq
so that, after integrating over
$k'_0$ one has, for $I^\epsilon({\vec x})$,
\beq
I^\epsilon({\vec x}) =
 - \frac{1}{8 \pi} 
\partial_i \int \frac{d^2k}{(2\pi)^2} \exp{i \vec{k}.\vec{x}} 
\left( \vert{\vec k} \vert K_1(\epsilon \vert {\vec k} \vert)  
- 4 m^2 \epsilon K_0(\epsilon \vert {\vec k} \vert)
+ \ldots \right)
\label{uch1}
\eeq
Using the modified Bessel functions expansion for small
$\epsilon$ one easily gets
\begin{eqnarray}
<[j_0({\vec x},t),j_i({\vec y},t]> & = &
\lim_{\epsilon \to 0}[-\frac{1}{8 \pi}\frac{1}{ \epsilon}
(1 + 4m^2 \epsilon^2 \log \epsilon)
 \times
 \partial_i\delta^{(2)}(\vec x - \vec y) + \nonumber\\
 & &   
 \frac{1}{16\pi} \epsilon \log \epsilon \partial_i \Delta 
 \delta^{(2)}(\vec x - \vec y)]
\label{sss}
\end{eqnarray}
The second and third terms in eq.(\ref{sss}), which vanish in the
equal-time limit, are precisely the $3$ dimensional analogs
of those encountered in $d=4$ dimensions \cite{Bra}-\cite{Ch}.  
Now, to compare our results with those
reported in \cite{dVG} using dimensional
regularization,  we 
shall expand $f(y)$ in eq.(\ref{exp7}) in
powers of $\epsilon$
(keeping the mass fixed) before integrating out $k'_0$. 
In this way, one ends with
\begin{eqnarray}
<[j_0({\vec x},t),j_i({\vec y},t]> & = & -\frac{1}{8 \pi}
\lim_{\epsilon \to 0}
\left(\frac{1}{ \epsilon}
 \partial_i\delta^{(2)}(\vec x - \vec y)  \right.\nonumber\\
 & & \left. -  \frac{\epsilon}{\Lambda} [4 m^2 
 \partial_i\delta^{(2)}(\vec x - \vec y)
 - \frac{1}{2}
   \partial_i \Delta 
 \delta^{(2)}(\vec x - \vec y)]\right)
\label{mnmnm}
\end{eqnarray}
where we have defined
\beq
\frac{1}{\Lambda} = \int_0^\infty dk'_0 \frac{1}{{k'}_0^2} \sin k'_0
\label{lam}
\eeq
In order to compare with ref.\cite{dVG} where current
commutators were computed using dimensional regularization,
we define, coming back to the original variable
$k_0 = k'_0/\epsilon$
\beq
A[d] = \frac{1}{2}
\int d^{d-2}k_0 \frac{1}{k_0^2} sin k_0 \epsilon
\label{dim}
\eeq
so that $A[d=3] = \epsilon/\Lambda$. One can now perform
the analytically continued integral to find, near $d=3$, the behavior
\beq
A[d] \sim - \epsilon \times \frac{\epsilon^{3-d}}{3-d}
\label{ambig}
\eeq
The same ambiguous result  eqs.(\ref{mnmnm}),(\ref{ambig})
for free fermions
is obtained in ref.\cite{dVG} near $d=3$. This ambiguity is however
removed by our previous procedure leading to eq.(\ref{sss}), the
pole in dimensional regularization corresponding as usual
to a logarithmic divergence. It is also interesting
to note that using in eq.(\ref{int1}) the nice approximation 
${\tilde F}_{appr}$ for ${\tilde F}$ proposed
in ref.\cite{BFO}, one can well check the correctness
of our previous analysis. In fact, on the one hand both procedures
leading to either eq.(\ref{sss}) or eq.(\ref{mnmnm}) give
(apart for the coefficient of $m^2$) the same result using 
${\tilde F}_{appr}$ or ${\tilde F}$. But, on the other hand,
for arbitrary mass, the the $k'_0$ integration can be easily
done {\it exactly} if one uses ${\tilde F}_{appr}$. The
subsequent $\epsilon \to 0$ limit gives precisely
(again apart for the coefficient of $m^2$) our result (\ref{sss})
for the equal-time current commutators.

From the analysis above, we see that not only the
infinite  Schwinger term, analogous to that
arising in  $d=4$ \cite{Bra}-\cite{Ch}
is obtained in the bosonized 
version of our $d=3$ fermion
theory but also the mass-dependent second term
as well as the triple derivative
third term in (\ref{sss}), both vanishing in
the equal time limit.

We then conclude that
 the equal-time commutator algebra for bosonic currents
$j_\mu = (1/\sqrt{4\pi}) \epsilon_{\mu \nu \alpha}
 \partial_\nu A_\alpha$
coincides with that for fermion currents,  
$ j_\mu = -i \bar\psi \gamma_{\mu} \psi$.
It is interesting to note at this point that
the $m \to 0$ theory
has the same  equal time current commutators that
the theory for arbitrary mass.  Now, for the massless
limit one obtains a very simple bosonized action.
Indeed, for $m \to 0$
  one  has from the expressions
(\ref{1.10})-(\ref{1.11}), (\ref{c1})-(\ref{c2})

\beq
{\tilde C}_1 = \frac{4 }{\pi k} \sin^2 \alpha
\label{em}
\eeq

\beq
{\tilde C}_2 = \frac{k}{\tan \alpha} {\tilde C}_1
\label{emi}
\eeq
\beq
{\tilde F} = \frac{1}{16k}
\label{efi}
\eeq
with $\tan \alpha = (\pi/4q)$ so that
 the bosonized action takes  the form \cite{BFO}
\beq
S_{bos} =  \frac{4}{\pi} \sin^2 \alpha\int d^3x [\frac{1}{4}
 F_{\mu \nu} 
\frac{1}{\sqrt{-\partial^2}} F_{\mu \nu} - 
\frac{i}{2} \cot \alpha \epsilon_{\mu \nu \lambda} 
A_\mu \partial_\nu A_\lambda]
\label{mar}
\eeq
which is nothing but the action found in \cite{Mar}
for free massless fermions. 
We thus see that our bosonized
effective theory correctly describes the commutator
algebra directly computed from the original fermionic theory.
This should be compared with the analysis in ref.\cite{Ban}
where the fermionic commutator algebra is  inferred from
the Maxwell-Chern-Simons algebra for electric and magnetic
fields using a bosonization recipe  which is  valid 
in the large mass limit. 
One can see that in the large mass 
regime,  terms depending on the product
$\epsilon m = \lambda $ will produce ambiguities according to the way
both limits ($\epsilon \to 0$ and $m \to \infty$) are taken
into account, a problem which is not present in the limit
of small masses.
To see this in more detail, let us come back to (\ref{int1})
and consider the case in which $\lambda$ is kept fixed while
$\epsilon \to 0$ (so that $m \to \infty$). In this case,
taking the limit before integrating
out $k'_0$,
one finds for $I^\epsilon$
\beq
I^\epsilon({\vec x}) \sim \vert m \vert h(\lambda)
 \partial_i \delta^{(2)}({\vec x})
\label{mg}
\eeq
where
\beq
h(\lambda) =
\frac{1}{2\pi} \int_0^\infty dz z\sin (2\lambda z) f(z)
\label{sui}
\eeq
with $f$ given by eq.(\ref{1.101}).
Let us note that using the approximate ${\tilde F}_{appr}$
of ref.\cite{BFO} and taking the limit 
after the exact integration over $k'_0$,
we recover the same behavior (\ref{mg}). 
We see that for  $\lambda = \epsilon m$ fixed, $h$ just gives 
a normalization factor so that one reproduces from $I^\epsilon$
in the form (\ref{mg}) a commutator algebra
at equal times and large mass that coincides with that
to be infered from
a Maxwell-Chern-Simons theory, namely (cf. eqs.(\ref{c21})-(\ref{b2}))
\beq
<[j_0({\vec x},t),j_i({\vec y},t]>
\longrightarrow 
c \vert m \vert \partial_i \delta^{(2)}({\vec x} - {\vec y}) 
~~~(m \to \infty)
\label{ulti}
\eeq
with $c$ a normalization constant. Again, currents appearing
in the l.h.s. of eq.(\ref{ulti}) are bosonic currents
which can be written  in terms of the electric and magnetic fields
thus reproducing  the MCS gauge invariant algebra \cite{Jac},\cite{DJ}.
One should note however that the free fermion - MCS mapping
is valid in the large mass limit of the original fermionic
theory, this meaning the large-distances regime for fermion fields.
Since current commutators test the short-distance regime, one
should not take the MCS gauge-invariant algebra as a starting
point to reproduce the fermion current commutators.

\section{Summary and conclusions}
As explained at length in the precedent sections, one can
establish in $d =3$ dimensional
space-time a bosonization recipe for fermion currents which is the
natural extension of the $2$ dimensional one. Namely, one has 
(apart from 
normalization factors)
\beq
-i \bar\psi \gamma_{\mu} \psi  \to \frac{1}{\sqrt{4\pi}}
\epsilon_{\mu \nu\alpha}
\partial_{\nu}A_{\alpha}, ~~~ (d=3)
\label{bosf3}
\eeq
to be compared with the well-known recipe
\beq
-i \bar\psi \gamma_{\mu} \psi  \to  \frac{1}{\sqrt{4\pi}}
\epsilon_{\mu \nu} \partial_{\nu} \phi, ~~~~ (d=2)
\label{bosf2}
\eeq
One can see eq.(\ref{bosf3}), which gives
an {\it exact} bosonization recipe, as the
natural extension to three dimensional space
of the  two-dimensional formula (\ref{bosf2}).
It is the bosonized action  that is much more complicated in
$3$ dimensions. Instead of a simple scalar action as in $2$ 
dimensions,
one has (in a quadratic approximation in the
auxiliary fields) 
\beq
\int  
\bar{\psi}(\id +  m )\psi d^{3}x \to  
\int[\frac{1}{4}F_{\mu\nu}
C_{1}(-\partial^2)F_{\mu\nu} - \frac{i}{2}A_{\mu}C_{2}(-\partial^2)
\epsilon_{\mu\nu\lambda}
\partial_{\nu}A_{\lambda}]d^3x  
\label{ac33}
\eeq
with $C_{1}$ and $C_{2}$ given by eqs.(\ref{rosa})-(\ref{1.11}).
Only in the limit of very large and very small fermion  mass the 
bosonized
action takes a simple form. As 
shown by eq.(\ref{mgran}) and in ref.\cite{BFO},  the bosonization 
relation (\ref{ac33})
becomes for large fermion mass the Maxwell-Chern-Simons
action \cite{FS}-\cite{FAS}
\begin{eqnarray}
\int  
\bar{\psi}(\id +  m )\psi d^{3}x 
& \longrightarrow &   
\int[\frac{1}{12 \vert m \vert}F_{\mu\nu}
F_{\mu\nu} \nonumber \\
& &\mp \frac{i}{2}A_{\mu}
\epsilon_{\mu\nu\lambda}
\partial_{\nu}A_{\lambda}] d^3x  ~~~ ~~ ~~
 ( m \to \infty )
\label{pac33}
\end{eqnarray}
On the other hand, for $m \to 0$
and in a quadratic approximation the action becomes
(see eq.(\ref{mar})) 
\begin{eqnarray}
\int  
\bar{\psi}(\id +  m )\psi d^{3}x 
& \longrightarrow & 
\frac{4}{\pi} \sin^2 \alpha\int  [\frac{1}{4} F_{\mu \nu} 
\frac{1}{\sqrt{-\partial^2}} F_{\mu \nu} \nonumber \\& &- 
\frac{i}{2} \cot \alpha  \epsilon_{\mu \nu 
\lambda} A_\mu \partial_\nu A_\lambda] d^3x ~~~ ~~ ~~
(  m \to 0 )
\label{marsi}
\end{eqnarray}
Either recipe (\ref{pac33}) or (\ref{marsi}) should be used according
to the regime one is to analyse. For example, in the study of 
Wilson loops in the long distance region, one employs 
relation (\ref{pac33}) finding an interesting connection
between the linking number and certain fermion loop operators 
\cite{FS}.
Concerning massless fermions one should use eq.(\ref{marsi})
which coincides (apart from normalization factors) to
that proposed 
in ref.\cite{Mar}, eq.(\ref{mar}).

In respect with current commutators, we have shown in this work 
that the bosonization recipe (\ref{bosf3}),(\ref{ac33}) reproduces
the equal time current commutator algebra corresponding to the
original  $3$-dimensional free massive
fermion model for arbitrary mass,
\beq
<[j_0({\vec x},t),j_i({\vec y},t]>  = -\frac{1}{8 \pi}
\lim_{\epsilon \to 0}
\frac{1}{ \epsilon}
\partial_i\delta^{(2)}(\vec x  -  \vec y)  
 \label{ult}
\eeq
with currents in the l.h.s. of eq.(\ref{ult}) expressed 
in terms of the bosonic field according to eq.(\ref{bosf3}).
As explained, this result can be obtained even if
one works in the $m \to 0$ limit and uses  bosonization recipes
(\ref{bosf3}),(\ref{marsi}) which implies a
simpler bosonic action. On the other hand,
the finite Schwinger term one finds starting 
directly from eq.(\ref{pac33})
is just the product of the way one takes the equal-time and the
$m \to \infty$ limits.  However, since the $m \to \infty$
limit explores the
large distances regime of the fermion theory, one should
not use the MCS gauge-invariant algebra to reproduce the
fermion current commutators which test short distances. 

In summary, the path-integral approach to bosonization has shown to
give a general bosonization recipe for free-fermions in $3$
space-time dimensions leading to the correct equal-time current
commutation algebra. This implies a step further in the obtention
of the fermion-boson mapping in $d=3$. As explained in refs.
\cite{FS},\cite{FS1}\cite{FAS}, this method is particularly apt to
treat interacting and non-Abelian extensions. Moreover, the
analysis of the $d >3$ case can be envisaged following a
similar approach. We hope to report on these
issues in a future work.

~

\underline{Acknowledgements}:
We would like to thank R. Jackiw for very useful
comments that prompted
this paper.
This work was
supported in part by  CICBA and CONICET, Argentina and the
Minist\`ere de l'Enseignement Sup\'erieure et de la
Recherche 
(France).
F.A.S.~thanks the Laboratoire de Physique Th\'eorique ENSLAPP
for its kind hospitality.

\end{document}